\documentclass[twocolumn,final]{elsart5p}

\bibstyle{elsart-num}

\usepackage{amssymb}

\journal{Physics Letters A}

\begin{document}
\begin{frontmatter}

\title{Exact\, stochastic\, Liouville\, and\, Schr\"{o}dinger\,
equations\, for\, open\, systems}
\author{Yuriy E. Kuzovlev}
\ead{\,\,\, kuzovlev@kinetic.ac.donetsk.ua
\,\,\,\,\,\,\,\,\,\,\,\,\,\,\,\,\,\,\,\,\,\,\,\,}
\address{Donetsk Institute for Physics and Technology of NASU,
ul.\,R.Luxemburg\,72\,, 83114 Donetsk, Ukraine}


\begin{abstract}
An universal form of kinetic equation for open systems is considered
which naturally unifies classical and quantum cases and allows to
extend concept of wave function to open quantum systems.
Corresponding stochastic Schr\"{o}dinger equation is derived and
illustrated by the example of inelastic scattering in quantum
conduction channel.
\end{abstract}


\end{frontmatter}

\section{Introduction}
It is seldom when one can say about a quantum system that it is in a
pure state described by certain wave function, $\,\Psi(t)\,$. Even
closed system generally is in a mixed state and should be described
by density matrix \cite{ll,bp,uw},
\begin{equation}
\begin{array}{c}
R(t)=\,\sum_{\alpha} |\Psi_{\alpha}(t)\rangle\,
P_{\alpha}\,\langle\Psi_{\alpha }(t)|\,\,\,,\label{expan}
\end{array}
\end{equation}
where $\,\Psi_{\alpha}(t)\,$ is a full set of mutually orthogonal
solutions to system's Schr\"{o}dinger equation
(\,$\,\langle\Psi_{\alpha}(t)|\Psi_{\beta}(t)\rangle=0\,$\, at
$\,\alpha\neq\beta\,$), time-independent $\,P_{\alpha}\,$\, are their
statistical weights, and whole $\,R(t)\,$ is solution to the von
Neumann equation. All the more if a system $\,\mathcal{S}\,$ under
interest is open, that is interacts with some other system
$\,\mathcal{W}\,$, e.g. ``the rest of the World''. In such case
representation (\ref{expan}) losses meaning and density matrix of
$\,\mathcal{S}\,$ becomes ``thing in itself'' to be found from von
Neumann equation of total closed system
$\,\mathcal{S}$+$\mathcal{W}\,$.

There are many different approaches to this vast problem
\cite{bp,uw,parth}. However, any of them results in either
approximate kinetic and stochastic equations for $\,\mathcal{S}\,$ or
formally exact but too complicated ones. In present paper, in part
following \cite{i1,i2,i3,i4,i5,i6}, one more approach is described
which leads to such kinetic equation which is at once simple and
exact. Its other advantages are intuitive obviousness, explicit
separation of contributions from $\,\mathcal{S}\,$ and
$\,\mathcal{W}\,$ and possibility to reformulate it in terms of wave
function of $\,\mathcal{S}\,$ itself, thus rehabilitating
representation (\ref{expan}) as applied to open systems.

\section{Stochastic treatment of dynamic interactions}
As usually, let us divide Hamiltonian $\,H\,$ of the whole system
$\,\mathcal{S}$+$\mathcal{W}\,$ into three parts and assume that the
interaction part, $\,H_{int}\,$, has bilinear form:
\begin{equation}
\begin{array}{c}
H\,=\,H_S+H_W+H_{int}\,\,\,,\,\,\,\,\,\, H_{int}\,=\,\sum_j
S_j\,W_j\,\,\,,\label{ham}
\end{array}
\end{equation}
where operators $\,S_j\,$ and $\,W_j\,$ act at different Hilbert
spaces of $\,\mathcal{S}\,$ and $\,\mathcal{W}\,$, respectively. Such
a decomposition of the interaction Hamiltonian is always possible
(and, as a rule, it directly follows from physics of interaction).
For any operator $\,\mathcal{O}\,$ define Liouville and Jordan
super-operators:
\[
\begin{array}{c}
\mathcal{L}(\mathcal{O})A\,\equiv\,\frac
i{\hbar}\,(\mathcal{O}A-A\mathcal{O})\,\,\,,
\,\,\,\,\,\,\Pi(\mathcal{O})A\,\equiv\,\frac 12
\,(\mathcal{O}A+A\mathcal{O})\,\,\,,
\end{array}
\]
where $\,A\,$ is arbitrary operator. Thus $\,\mathcal{L}(H)\,$ is the
Liouville evolution operator of $\,\mathcal{S}$+$\mathcal{W}\,$. One
can see that correspondingly to (\ref{ham}) it decomposes to
\begin{eqnarray}
\mathcal{L}(H)\,=\,\mathcal{L}(\,H_S+H_W+\sum S_jW_j\,
)\,=\,\,\,\,\,\,\,\,\,\,\,\,\,\,\,\,\,\,\,\,\,\,\,\,\label{l}
\\=\,\mathcal{L}(H_S)+\mathcal{L}(H_W)+\sum\,[\,
\mathcal{L}(S_j)\,\Pi(W_j)+\Pi(S_j)\,\mathcal{L}(W_j)\,]\nonumber
\end{eqnarray}
From the viewpoint of $\,\mathcal{S}\,$ it is formally equivalent to
\begin{eqnarray}
\mathcal{L}_S(t)\,\equiv\,\mathcal{L}(H_S)+\sum
\,[\,x_j(t)\,\mathcal{L}(S_j)\,+\, y_j(t)\,\Pi(S_j)\,]\,\,\label{xys}
\end{eqnarray}
where $\,x_j(t)\,$ and $\,y_j(t)\,$ play as arbitrary scalar time
functions although in respect to $\,\mathcal{W}\,$ they are
super-operators
\[
\begin{array}{c}
x_j(t)\,=\, e^{-\,\mathcal{L}(H_W)\,t}\,\Pi(W_j)\,
e^{\,\mathcal{L}(H_W)\,t}\,\,\,,
\end{array}
\]
\[
\begin{array}{c}
y_j(t)\,=\, e^{-\,\mathcal{L}(H_W)\,t}\,\mathcal{L}(W_j)\,
e^{\,\mathcal{L}(H_W)\,t}\,\,
\end{array}
\]
In detail, if\, $\,\rho(t)\,$\, denotes joint density matrix of
$\,\mathcal{S}$+$\mathcal{W}\,$, which undergoes the von Neumann
equation $\,\dot\rho =\mathcal{L}(H)\rho\,$,\, and
$\,\rho_S(t)=\,$\,Tr$_{\,W}\rho(t)\,$\, density matrix of
$\,\mathcal{S}\,$,\, then
\begin{eqnarray}
\rho_S(t)\,=\,\texttt{Tr}_{\,W}\,\,\overleftarrow{\exp
}\left[\int^t_0 \mathcal{L}_S(t^{\,\prime\,})
\,dt^{\,\prime}\right]\rho(0)\,\,\,,\label{rs}
\end{eqnarray}
where $\,\overleftarrow{\exp }\,$ means chronologically ordered
exponential.

These facts prompt us to consider $\,x_j(t)\,$ and $\,y_j(t)\,$ as
random processes while operation $\,\texttt{Tr}_{\,W}\,$ in
(\ref{rs}) as statistical averaging over them. In order to realize
such ``stochastic representation of dynamic interaction'' between
$\,\mathcal{S}\,$ and $\,\mathcal{W}\,$ in a simple unambiguous way,
let us assume that at some initial time moment $\,t_0\,$, e.g.
$\,t_0=0\,$, before the interaction $\,\mathcal{S}\,$ and
$\,\mathcal{W}\,$ were statistically independent one on another:\,
$\,\rho(t_0)=\rho_{S}^{(in)}\rho_W^{(in)}\,$. Then we can write
\begin{eqnarray}
\rho_S(t)\,=\,\langle\,R(t)\,\rangle\,\,\,,\label{sr}
\end{eqnarray}
where $\,R(t)\,$ satisfies ``stochastic Lioville equation''
\begin{eqnarray}
\frac {d R(t)}{dt}\,=\,\mathcal{L}_S(t)\,R(t)\,\,\label{sle}
\end{eqnarray}
with initial condition $\,R(t_0)=\rho_{S}^{(in)}\,$, and angle
brackets $\,\langle\,...\,\rangle\,$ designate the averaging. At
that, evidently,  for any statistical moment\,\,
$\,\langle\,...\,\rangle
=\,$Tr$_{\,W}\,\,\overleftarrow{T}\,...\,\,\rho_W^{(in)}\,$,\,\,
where\, $\,\overleftarrow{T}\,$\, symbolizes chronological
ordering,\, and complete statistics of random processes $\,x_j(t)\,$
and $\,y_j(t)\,$ is summarized by their characteristic functional as
follows,
\begin{equation}
\begin{array}{c}
\langle \,\,\exp \int \sum\,[\,u_j(t)\,x_j(t)+f_j(t)\,
y_j(t)\,]\,\,dt\,\,\rangle\,\,=\label{cf0}\\
=\,\texttt{Tr}_{\,W}\,\,\overleftarrow{\exp }\left[\int
\mathcal{L}_W(t)\,dt\right]\rho_W^{(in)}\,\,\,,\\
\mathcal{L}_W(t)\,\equiv\,\mathcal{L}(H_W)+\sum\,[\,u_j(t)\,
\Pi(W_j)+f_j(t)\,\mathcal{L}(W_j)\,]
\end{array}
\end{equation}
Here $\,u_j(t)\,$ and $\,f_j(t)\,$ are arbitrary probe functions, the
integrations begin at $\,t=t_0\,$, and, if necessary, $\,t_0\,$ can
be moved away to $\,-\,\infty\,$.

Physical meaning of $\,x_j(t)\,$ in (\ref{xys}) and
(\ref{rs})-(\ref{sle}) is quite clear from (\ref{xys}):\,
$\,x_j(t)\,$ merely reproduce Hamiltonian perturbation of
$\,\mathcal{S}\,$ by $\,\mathcal{W}\,$ as prescribed by (\ref{ham}).
In order to understand meaning of $\,y_j(t)\,$ there, notice that
according to (\ref{cf0}) the super-operators $\,\Pi(W_j)\,$ represent
observation of variables $\,W_j\,$ in subsystem $\,\mathcal{W}\,$ by
subsystem $\,\mathcal{S}\,$. Hence, analogously and symmetrically,
super-operators $\,\Pi(S_j)\,$ in $\,\mathcal{L}_S(t)\,$ correspond
to observation of variables $\,S_j\,$ in $\,\mathcal{S}\,$ by
$\,\mathcal{W}\,$ while $\,y_j(t)\,$, according to (\ref{cf0}),
represent conjugated Hamiltonian perturbation of $\,\mathcal{W}\,$ by
$\,\mathcal{S}\,$. Therefore $\,y_j(t)\,$ introduce also response of
$\,\mathcal{W}\,$ to this perturbation and eventually various
feedback effects, in particular, dissipation and renormalization of
dynamic properties of $\,\mathcal{S}\,$ .

By these reasons, generally $\,x_j(t)\,$ in (\ref{sle}) behave like
conventional random processes, but $\,y_j(t)\,$ are extremely unusual
ones. Indeed, if $\,u_j(t)=0\,$ in (\ref{cf0}) then
$\,\mathcal{L}_W(t)\,$ turns into purely Liouville super-operator,
therefore the characteristic functional turns into unit regardless of
$\,f_j(t)\,$. Thus any statistical moment of $\,y_j(t)\,$ themselves
equals to zero, $\,\langle y(t_1)...y(t_n)\rangle =0\,$. Moreover,
any moment where some $\,y_j(t)\,$ has most late time argument also
is zero: $\,\langle x(\tau_1)...\,x(\tau_m)\,y(t_1)...\,y(t_n)\rangle
=0\,$ if $\,\max_j(t_j)\geq \max_j(\tau_j)\,$.\, However, in general
$\,\langle\, x(\tau)\,y(t<\tau)\,\rangle \neq 0\,$. Physically, these
features of $\,y_j(t)\,$ trivially mean that perturbation of
$\,\mathcal{W}\,$ by $\,\mathcal{S}\,$ (and conjugated observation of
$\,\mathcal{S}\,$ by $\,\mathcal{W}\,$) always foregoes response of
$\,\mathcal{W}\,$ to it, in agreement with the causality principle.
At that, $\,\langle\, x(\tau)\,y(t<\tau)\,\rangle \,$ and other
cross-correlations just represent dissipation in $\,\mathcal{S}\,$.

In the classical limit super-operators $\,\mathcal{L}(...)\,$ and
$\,\Pi(...)\,$ become the Poisson bracket and mere multiplication by
a phase function, respectively, but physical sense of $\,x(t)\,$ and
$\,y(t)\,$ and their main statistical properties do not change. More
details about them and various generalizations of the ``stochastic
representation'' can be found in \cite{i1,i3,i4,i6}.

\section{Stochastic Schr\"{o}dinger equation}
If initially (before the interaction) $\,\mathcal{S}\,$ was in a pure
state, $\,\rho_{S}^{(in)}=|\Psi^{(in)}\rangle\langle \Psi^{(in)}|\,$,
then equation (\ref{sle}) has a solution which all the time keeps
such form:\, $\,R(t)=|\Psi(t)\rangle\langle \Psi(t)|\,$,\, where
$\,\Psi(t)\,$ satisfies stochastic Schr\"{o}dinger equation
\begin{equation}
\frac {d \Psi(t)}{dt}\, =-\frac i{\hbar}\,[\,H_S+\sum\,w_j(t)\,
S_j\,]\,\Psi(t)\,\,\label{sse}
\end{equation}
with initial condition\, $\,\Psi(t_0)=\Psi^{(in)}\,$, and
\[
\begin{array}{c}
w_j(t)\,\equiv\,x_j(t)\,+\,i\hbar\, y_j(t)/2\,
\end{array}
\]
are mentioned as complex-valued random processes. Hence, solution to
(\ref{sle}) under arbitrary initial condition $\,R(t_0)\,$, let\,
$\,\rho_{S}^{(in)}=\sum_{\alpha }\,|\Psi^{(in)}_{\alpha
}\rangle\,P_{\alpha }\,\langle \Psi^{(in)}_{\alpha }|\,$\, in the
diagonal form, can be written in the form (\ref{expan}) where each of
$\,\Psi_{\alpha }(t)\,$ evolves according to (\ref{sse}) but all
$\,P_{\alpha }\,$ stay constant.

Because of complexity of $\,w_j(t)$, the evolution governed by
(\ref{sse}) is not unitary:
\begin{equation}
\frac {d}{dt}\,\langle\Psi_{\alpha}(t)|\Psi_{\beta}(t)\rangle
\,=\,\sum\,
y_j(t)\,\langle\Psi_{\alpha}(t)|S_j|\Psi_{\beta}(t)\rangle\,\neq\,
0\,\ \label{nu}
\end{equation}
Nevertheless it is unitary on average:
\begin{eqnarray}
\frac {d}{dt}\,\left\langle\,
\langle\Psi_{\alpha}(t)|\Psi_{\beta}(t)\rangle\,\right\rangle
=\langle\sum\,
y_j(t)\langle\Psi_{\alpha}(t)|S_j|\Psi_{\beta}(t)\rangle\,\rangle
= 0\,\,, \nonumber\\
\left\langle\,\,\langle\Psi_{\alpha}(t)|\Psi_{\beta}(t)\rangle\,
\,\right\rangle =
\langle\Psi_{\alpha}(t_0)|\Psi_{\beta}(t_0)\rangle\,\propto\,
\delta_{\alpha\beta} \,\,\,\,\,\,\,\,\,\,\,\,\,\,\,\,\,\, \label{au}
\end{eqnarray}
This statement directly follows from the above mentioned specific
statistical properties of $\,y_j(t)\,$.

Let the interaction by its nature is time-localized, that is has a
character of scattering. In such situation for any particular
``stochastic pure state'' one can write\,
\[
\begin{array}{c}
|\Psi(t)\rangle\, =\, |\Psi^0(t)\rangle\,+\,|\Psi^s(t)\rangle\,\,
\end{array}
\]
with $\,|\Psi^0(t)\rangle \,$ describing free evolution of
$\,\mathcal{S}\,$ and $\,|\Psi^s(t)\rangle\,$ at
$\,t\rightarrow\infty\,$ representing result of the scattering. Then
due to (\ref{au})\, $\,\langle\,\langle\Psi(t)|\Psi(t)\rangle
\,\rangle =\langle\Psi^0|\Psi^0\rangle \,$ and consequently
\begin{equation}
\begin{array}{c}
2\,\texttt{Re}\, \langle\,\Psi^{0}|\langle
\,\Psi^s\,\rangle\rangle\,+\, \langle
\,\langle\,\Psi^s|\Psi^s\rangle\,\rangle \,=\,0\,\,\label{sau}
\end{array}
\end{equation}
This is nothing but ``the optical theorem on average'' which equally
holds for both elastic and inelastic scattering. It demonstrates also
that average wave function $\,\langle \,\Psi(t)\,\rangle\,$ can bring
useful information about an open quantum system.

\section{Interaction with thermal bath}
Of special interest are cases when $\,\mathcal{W}\,$ is very large
system in thermodynamical equilibrium,\,
$\,\rho_{W}^{(in)}\,\propto\,\exp{(-\,H_W/T)}\,$\,, so that all
$\,x_j(t)\,$, $\,y_j(t)\,$ and $\,w_j(t)\,$ are stationary random
processes. If one assumes (without loss of generality) that
$\,\texttt{Tr}_W\,W_j\,\rho_{W}^{(in)}\,=0\,$, then all the processes
have zero mean values and all their pair correlations reduce to
function
\[
\begin{array}{c}
K_{jm}(\tau )\,\equiv\, \texttt{Tr}_W\,\,W_j\,\exp{(-i\tau H_W/\hbar
)}\,W_m\,\rho_{W}^{(in)}\,
\end{array}
\]
One can routinely verify that
\begin{equation}
\begin{array}{c}
K_{jm}(\tau )\, =\,K_{mj}^{*}(-\tau)\,\,\,,\\
\langle w^{*}_j(\tau )\, w _m(0)\rangle \,=\,K_{jm}(\tau )\,\,\,,\\
\langle w _j(\tau )\,w _m(0)\rangle \,=\,\langle w^{*}_j(\tau
)\,w^{*}_m(0)\rangle ^{*}\,=\,\\
=\,K_{jm}(\tau )\, \theta(\tau)+K_{jm}^{*}(\tau ) \,\theta(-\tau)\,\,\,,\\
K_{jm}^{xx}(\tau )\,\equiv\, \langle x_j(\tau )\,x_m(0)\rangle
\,=\,\texttt{Re} \,K_{jm}(\tau )\,\,,\label{rcf}\\
K_{jm}^{xy}(\tau )\,\equiv\, \langle x_j(\tau )\,y_m(0)\rangle
\,=\,(2/\hbar)\,\theta(\tau)\,\texttt{Im}\,\,K_{jm}(\tau )\,\,\,,\\
K_{jm}(\tau -i\hbar /2T)\,=\,K_{mj}(-\tau-i\hbar /2T)\,\,\,,
\end{array}
\end{equation}
where $\,\theta(t)\,$ is the Heaviside step function. The last
equality expresses equilibrium of $\,\mathcal{W}\,$ and in its turn
implies definite relation between $\,K_{jm}^{xx}(\tau )\,$ and
$\,K_{jm}^{xy}(\tau )\,$:
\begin{eqnarray}
K_{jm}^{xx}(\tau )=\int_0^\infty \cos (\omega \tau )\,\sigma
_{jm}(\omega )d\omega \,\,\,,\,\,\,\,\,\,\,\,\,\,\,\label{k}\\
K_{jm}^{xy}(\tau )=-\frac {2\theta (\tau)}{\hbar}\int_0^\infty \sin
(\omega \tau )\tanh \left[\frac {\hbar\omega}{2T}\right] \sigma
_{jm}(\omega )\,d\omega\,\,\,,\nonumber
\end{eqnarray}
where $\,\sigma _{jm}(w )\,$ is non-negatively defined spectrum
matrix. In essence, this is usual fluctuation-dissipation theorem.

Of course, the $\,\mathcal{W}\,$'s equilibrium results also in
definite relations between higher-order correlators of $\,x(t)\,$ and
$\,y(t)\,$. They were discussed in \cite{i7}. In the classical limit
all they can be summarized by symbolic equality\,
\[
\varepsilon_j\,y_j(-t)\,\asymp\,y_j(t)\,+\,\frac 1T\,\frac {d
x_j(t)}{dt}\,\,\,,
\]
where $\,\asymp\, $ means statistical equivalence under time
inversion, $\,t\rightarrow-t\,$, and $\,\varepsilon_j\,\pm 1\,$\,
indicates time parity of variable $\,x_j(t)\,$ from viewpoint of
Hamiltonian mechanics, so that $\,x_j(-t)\asymp \varepsilon_j
x_j(t)\,$. Correspondingly, the second-order relation expressed by
(\ref{k}) in the classical limit simplifies to
\[
K_{jm}^{xy}(\tau )\,=\, \frac {\theta(\tau )}{T}\,\frac {d}{d\tau }\,
K_{jm}^{xx}(\tau )
\]

\section{Inelastic scattering center in conducting channel}
Let $\,\mathcal{S}\,$ be one-dimensional conduction channel formed by
discrete sites ($\,n=...-1,0,1,...\,$), $\,\mathcal{W}\,$ be
thermostat, and their interaction realizes through randomly varying
potential localized at only one site, for certainty, $\,n=0\,$. Then
$\,(H_S)_{n^{\prime}n}=\epsilon\,
[\,2\delta_{n^{\prime}n}-\delta_{n^{\prime}n-1}
-\delta_{n^{\prime}n+1}\,]/2\,$, where $\,\epsilon\,$ is channel's
bandwidth, and $\,\sum w_j(t)S_j= w(t)\,S\,$, where
$\,S_{n^{\prime}n}=\delta_{n0}\delta_{n^{\prime}0}\,$. We want to
consider evolution of wave function $\,\Psi(t,n)\,$ which initially,
at $\,t=t_0<0\,$, represents a wide wave packet placed on the left
from $\,n=0\,$ and moving to the right with momentum $\,\hbar
k_0>0\,$. Thus let us divide it into free and scattered parts:
$\,\Psi(t,n)=\Psi^{0}(t,n)+\Psi^s(t,n)\,$, where
$\,\dot\Psi^{0}=-\,iH_S\,\Psi^{0}/\hbar\,$, so that (\ref{sse})
transforms to
\begin{equation}
\begin{array}{c}
\frac {d \Psi^s(t,n)}{dt}=-\frac i{\hbar}\,[(H_S\Psi^s)(t,n)+
w(t)\,\delta_{n0}\,\Psi(t,0) \,] \label{sse1}\,\,\,,
\end{array}
\end{equation}
with initial condition $\,\Psi^s(t_0,n)=0\,$.

Strictly speaking, such division presumes that particle scattered by
fluctuating potential can not be captured by it. If it is really so,
that is $\,\Psi^s(t,n)\rightarrow 0\,$ at $\,t\rightarrow\infty\,$
for any given finite $\,n\,$ (first of all, $\,n=0\,$), then at large
enough $\,t>0\,$ after scattering we can determine probabilities of
reflection, $\,\Re\,$, and transmission, $\,\mathcal{T}\,$, as
\begin{equation}
\begin{array}{c}
\mathcal{T}\,=\,\sum_{n\geq\,0}|\Psi|^2\,\,\,,
\,\,\,\,\,\,\,\,\,\,\,\,\,\,\,\,\,\,\,\,\,\,\,\,\,\,\label{re}\\
\Re =\,\sum_{n<\,0}|\Psi |^2\, = \sum_{n<\,0}|\Psi^s|^2\,= \,\frac 12
\sum_{n} |\Psi^s|^2\,
\end{array}
\end{equation}
Here we took into account mirror symmetry of scattered wave,
$\,\Psi^s(t,-n)=\Psi^s(t,n)\,$, which obviously follows from
(\ref{sse1}). After averaging (\ref{re}) over the thermostat and
combining with general ``optical theorem'' (\ref{sau}) we have
\begin{equation}
\begin{array}{c}
\langle\,\Re\,\rangle \,=\,1-\langle\,\mathcal{T}\,\rangle\,=\,
-\,\texttt{Re}\,\sum_n\, \Psi^{0*}(t,n)\,\langle\Psi^s(t,n)\rangle\,
\label{mre}
\end{array}
\end{equation}
Alternatively, we can describe the scattering in terms of outgoing
wide wave packets with discrete wave numbers $\,k\,$ (separated by a
width of ingoing packet in reciprocal space) and amplitudes
$\,\Psi_k=\Psi_k^0+\Psi_k^s\,$. Then $\,\Psi_k^0=\delta_{kk_0}\,$,
and optical theorem (\ref{mre}) takes form
\begin{equation}
\langle \Re \rangle\,=\,-\,\texttt{Re\,}\, \langle
\Psi_{k_0}^s\rangle\, =\,-\,\texttt{Re\,}\,\langle
\Psi_{-\,k_0}^s\rangle \,\,\label{ot}
\end{equation}
The second equality here is again consequence of the mirror symmetry
of scattering which implies\, $\,\Psi_{-\,k}^s=\Psi_{k}^s\,$\,.

Next, let us divide $\,\Re \,$ and $\,\mathcal{T} \,$ into elastic
and inelastic parts marked by ``el'' and ``in'', respectively, so
that
\begin{equation}
\begin{array}{c}
\Re^{\,el}\, =\,|\Psi_{-k_0}|^2\,\,,\,\,\,\,\,
\mathcal{T}^{\,el}\,=\,  |\Psi_{k_0}|^2\,\,\,,\label{elin}
\end{array}
\end{equation}
and use obvious general inequalities
\[
\begin{array}{c}
\langle|\Psi_{k}|^2\rangle\, \geq\, |\langle
\Psi_{k}\rangle|^2\,\geq\, (\texttt{Re\,}\,\langle
\Psi_{k}\rangle)^2\,
\end{array}
\]
By applying them to (\ref{elin}) and combining with (\ref{ot}) it is
easy to obtain
\begin{equation}
\begin{array}{c}
\langle\Re^{\,in}\rangle = \langle\Re\rangle-
\langle\Re^{\,el}\rangle \leq
\langle\Re\rangle-\langle\Re\rangle ^2\,\,,\\
\langle\mathcal{T}^{\,in}\rangle =\langle\mathcal{T}\rangle-
\langle\mathcal{T}^{\,el}\rangle \leq
\langle\mathcal{T}\rangle-\langle\mathcal{T}\rangle ^2
\,\,\,\,\label{ins}
\end{array}
\end{equation}
In fact these are identical inequalities since $\,\Re^{\,in}
=\mathcal{T}^{\,in}\,$ already due to the mirror symmetry. Their sum
gives us restriction on total probability of inelastic scattering,
$\,\mathcal{P}^{in}\,$\,:
\begin{equation}
\begin{array}{c}
\mathcal{P}^{in}\,=\,\langle\Re^{\,in}\rangle
+\langle\mathcal{T}^{\,in}\rangle \,\leq\,
\,2\,\langle\Re\rangle\,(1-\langle\Re\rangle )\,\leq\, 1/2
\,\label{restr}
\end{array}
\end{equation}
This seems interesting and important consequence from the ``unitarity
on average''.

More careful consideration bases on stochastic integral equation for
$\,\psi (t)\equiv \Psi(t,0)\,$ which follows from (\ref{sse1}):
\begin{equation}
\begin{array}{c}
\psi(t)\,=\,\Psi^{0}(t,0)-\frac
i{\hbar}\int_{t_0}^{t}G_0(t-t^{\prime},0)\,w(t^{\prime})
\,\psi(t^{\prime})\,dt^{\prime}\,\,\,, \label{ise}\\
G_0(\tau,n)\,\equiv\,\int_{-\pi}^{\pi}\exp
\left[\,ink-\,iE(k)\tau/\hbar \,\right]\,\frac {dk}{2\pi}\,\,\,,
\end{array}
\end{equation}
with $\,E(k)=\epsilon\, [\,1-\cos\,{k}\,]/2\,$\, being dispersion of
the channel. Let us confine ourselves by weak scattering, in the
sense that $\,\langle w(t)\rangle =0\,$ and
$\,\sigma(\omega)\,\ll\,\hbar\,\epsilon\,$\,, where spectrum
$\,\sigma(\omega)\,$ of the fluctuating potential is mentioned like
in (\ref{k}) as spectrum of its real part $\,x(t)=\texttt{Re\,}w(t)$:
~$\,\langle\,x(\tau),x(0)\rangle\,=\int_0^{\infty}\,\cos(\omega\tau)\,
\sigma(\omega)\,d\omega$\,.\, Thus, according to
(\ref{rcf})-(\ref{k}),
\begin{equation}
\begin{array}{c}
\langle\,w(\tau)\,w(0)\rangle\,=\,K(|\tau |)\,\,\,,\,\,\,\,\,
\langle\,w^{*}(\tau)\,w(0)\rangle\,=\,K(\tau )\,\,\,,\label{kk}\\
K(\tau)\,=\,\int_0^{\infty}\frac
{e^{i\omega\tau}+\exp(\hbar\omega/T)e^{-\,i\omega\tau}}
{1+\exp(\hbar\omega/T)}\,\,\sigma(\omega)\,d\omega\,\,
\end{array}
\end{equation}
Then simple analysis of equation (\ref{ise}) in the corresponding
standard ``one-loop'' approximation (or, in other words, ``the
Bourret approximation'' \cite{kl}) yields
\begin{equation}
\begin{array}{c}
\langle\Re\rangle \,=\,\frac
{C}{1+C}\,\,\,,\,\,\,\,\,\,\,C\,\approx\, \frac
{1}{\sqrt{E_0(\varepsilon -E_0)}}\,\,\times \,\label{ap}
\end{array}
\end{equation}
\[
\begin{array}{c}
\times \int_{-E_0/\hbar}^{(\epsilon -E_0)/\hbar}\frac
{\sigma(|\omega|)\,d\omega
}{[1+\exp{(\hbar\omega/T)}]\,\sqrt{(E_0+\hbar\omega)(\varepsilon
-E_0-\hbar\omega)}} \,\,\,,
\end{array}
\]
where $\,E_0\equiv E(k_0)\,$. In fact, due to our assumption, $\,C\ll
1\,$ and at that, naturally, almost all the scattering is inelastic,
except may be case of very small particle's energy $\,E_0\,$ (for
more details see \cite{i5}). Contributions to the integral from
$\,\omega >0\,$ and $\,\omega <0\,$ correspond to scattering with
energy radiation or absorption by the thermostat, respectively.

\section{Conclusion}
We just had a chance to make sure that there exists universal form of
stochastic Liouville and Schr\"{o}dinger equations which is exactly
valid for arbitrary open quantum system $\,\mathcal{S}\,$
representable as a part of some closed system\,
$\,\mathcal{S}$+$\mathcal{W}\,$.\, Simultaneously there exist
universal exact recipes for determination of statistical
characteristics of random sources, or noises, $\,w(t)\,$, what enter
these equations.

Peculiarity of thus stated ``stochastic representation of dynamic
interactions'' (between $\,\mathcal{S}\,$ and $\,\mathcal{W}\,$) is
that stochastic images, $\,w(t)\,$, of Hermitian operator variables
$\,W(t)\,$ of $\,\mathcal{W}\,$ behave like a $\,c$-number (purely
commutative) complex-valued random processes. Their imaginary parts,
$\,y(t)=\texttt{Im\,}\,\,w(t)\,$, introduce dissipation into
$\,\mathcal{S}$'s evolution by means of virtual violation of its
unitarity. Nevertheless, due to specific statistical properties of
$\,y(t)\,$, statistical averaging restores the unitarity although
keeps the dissipation.

It is useful to underline that our approach to open systems makes it
possible to obtain final results under interest without
conventionally thought ``kinetic equations'', i.e. omit stages of
approximate derivation of a kinetic equation and then its approximate
solving. Illustrations were done in \cite{i1,i2,i3,i4,i5} and just
above. From the other hand, our approach gives all tools for correct
construction of kinetic equations as well as ``Langevin equations''
for $\,\mathcal{S}$'s variables \cite{i6}.

Importantly, Hamiltonian character of joint dynamics of
$\,\mathcal{S}$+$\mathcal{W}\,$ implies general statistical
connections between real and imaginary components of $\,w(t)\,$
(especially when $\,\mathcal{W}\,$ is a thermostat \cite{i7}), which
help to convert simple abstract models of stochastic processes into
our approach. At that, one can exploit all the theory of linear
dynamic systems with randomly varying parameters (including theory of
waves in random media \cite{kl}).



\end{document}